\documentclass[10pt,conference]{IEEEtran}\IEEEoverridecommandlockouts
\usepackage{epsfig,graphicx,subfigure,psfrag,amsmath,cases}
\usepackage{latexsym,amssymb,amsmath,epsfig,subfigure,algorithm,mathtools}
\usepackage{algorithmic}
\usepackage{color}
\usepackage{url}
\usepackage{scrtime}
\usepackage[left=0.6in,top=0.43in,bottom=0.43in,right=0.6in]{geometry}
\author{\authorblockN{Yan Sun\authorrefmark{1}, Derrick Wing Kwan
Ng\authorrefmark{2}, Nikola Zlatanov\authorrefmark{3}, and Robert Schober\authorrefmark{1}
\thanks{Robert Schober is also with the University of British
Columbia, Canada. }
}\vspace*{-0mm}
Institute for Digital Communications, Friedrich-Alexander-University
Erlangen-N\"urnberg (FAU), Germany\authorrefmark{1}\\
School of Electrical Engineering and Telecommunications, The University
of New South Wales, Australia\authorrefmark{2} \\
Department of Electrical and Computer Systems Engineering, Monash University, Australia\authorrefmark{3} \vspace*{-3mm}
}

\title{Robust Resource Allocation for Full-Duplex Cognitive Radio Systems}

\date{\thistime,\,\today}

\newtheorem{Thm}{Theorem}
\newtheorem{Lem}{Lemma}

\newtheorem{T-Prob}{Transformed Problem}

\DeclareMathOperator{\Tr}{Tr}
\DeclareMathOperator{\Rank}{Rank}
\DeclareMathOperator{\maxo}{maximize}
\DeclareMathOperator{\mino}{minimize}
\DeclareMathOperator{\diag}{\mathrm{diag}}
 \newcommand{\qed}{\hfill \ensuremath{\blacksquare}}

\newcommand{\abs}[1]{\lvert#1\rvert}
\newcommand{\norm}[1]{\lVert#1\rVert}
\begin{document}
\IEEEspecialpapernotice{(Invited Paper)}
\maketitle
\begin{abstract}
In this paper, we investigate resource allocation algorithm design for full-duplex (FD) cognitive radio systems. The secondary network employs a  FD base station for serving multiple half-duplex downlink and uplink users simultaneously. We study the resource allocation design for minimizing the maximum interference leakage to primary users while providing quality of service for secondary users. The imperfectness of the channel state information of the primary users is taken into account for robust resource allocation algorithm design. The algorithm design is formulated as a non-convex optimization problem and solved optimally by applying semidefinite programming (SDP) relaxation. Simulation results not only show the significant reduction in interference leakage compared to baseline schemes, but also confirm the robustness of the proposed algorithm.
\end{abstract}
%\begin{keywords}
%\end{keywords}
\vspace*{-0mm}
\section{Introduction}
\label{sect1}
Bandwidth has become a scarce resource in wireless systems due to the tremendous demand for ubiquitous and high data rate communication.
Recently, cognitive radio (CR) has emerged as a promising paradigm to improve spectrum efficiency. In particular, CR technology allows a secondary network to share the spectrum of a primary network without severely degrading the quality of service (QoS) of the primary network.
The authors of \cite{kwan2015layered} proposed an optimal beamforming and power control algorithm to guarantee communication security in multiuser CR networks.
In \cite{tajer2010beamforming}, distributed beamforming and rate allocation for multiple secondary users were considered for maximization of the minimum data rate achieved by secondary users.
However, the spectral resource is still underutilized in \cite{kwan2015layered,tajer2010beamforming}. Specifically, since the secondary network operates in the traditional half-duplex (HD) mode, orthogonal radio resources are used for uplink (UL) and downlink (DL) transmission which limits the spectral efficiency.

Full-duplex (FD) wireless communication has recently attracted significant research interest due to its potential to double the spectral efficiency by performing simultaneous DL and UL transmission using the same frequency \cite{YanFD2016Journal,Kwan2016FDDAS}. Therefore, it is expected that the spectral efficiency of existing wireless communication systems can be further improved by employing an FD base station (BS) in CR networks.
However, the simultaneous UL and DL transmission may lead to excessive interference leakage to the primary network and degrade the quality of communication.
Therefore, different resource allocation designs for FD-CR networks were proposed to overcome this challenge.
For example, the authors of \cite{zheng2013full} studied the rate region of a secondary single-antenna user served by a secondary FD BS while guaranteeing the primary user's QoS.
In \cite{afifi2015incorpor}, a suboptimal resource allocation algorithm was proposed for the maximization of the sum throughput of secondary FD users.
% In \cite{cirik2015mse}, a joint transmit- and receive-beamforming was designed to minimize the sum of mean-squared errors (MSE) of secondary users in a FD CR system.
However, \cite{zheng2013full,afifi2015incorpor}
%\nocite{afifi2015incorpor}--\cite{cirik2015mse}
assumed that the channel state information (CSI) of the link between the secondary network and the primary network is perfectly known at the secondary FD BS which is a highly idealistic assumption. In fact, the perfect CSI of the primary users may not be available at the secondary FD BS since they do not directly interact with the secondary network.
Besides, the objective of the resource allocation algorithms in \cite{zheng2013full,afifi2015incorpor} was to improve the performance of the secondary network from the secondary network's point of view. However, in FD-CR systems, interference leakage is more serious than in traditional HD-CR systems due to the simultaneous secondary DL and UL transmission. Therefore, in FD-CR systems, a careful design of the resource allocation is necessary.

Motivated by the aforementioned observations, we formulate an optimization problem to minimize the maximum interference leakage caused by the secondary FD network to the primary network while guaranteeing the QoS of all secondary users. The imperfectness of the CSI of the interference leakage channels is taken into account in the proposed problem formulation to facilitate a robust resource allocation.

\section{System Model}
In this section, we present the considered FD-CR wireless communication system model.
\vspace*{-0mm}
\subsection{Notation}%
We use boldface capital and lower case letters to denote matrices and vectors, respectively. $\mathbf{A}^H$, $\Tr(\mathbf{A})$, and  $\Rank(\mathbf{A})$ denote the Hermitian transpose, trace, and rank of matrix $\mathbf{A}$, respectively; $\mathbf{A}\succeq \mathbf{0}$ and $\mathbf{A}\succ \mathbf{0}$ indicate that $\mathbf{A}$ is a positive semidefinite and a positive definite matrix, respectively; $\mathbf{I}_N$ is the $N\times N$ identity matrix; $\mathbb{C}^{N\times M}$ denotes the set of all $N\times M$ matrices with complex entries; $\mathbb{H}^N$ denotes the set of all $N\times N$ Hermitian matrices; $\abs{\cdot}$ and $\norm{\cdot}$ denote the absolute value of a complex scalar and the Euclidean vector norm, respectively; ${\cal E}\{\cdot\}$ denotes statistical expectation; $\diag(x_1, \cdots, x_K)$ denotes a diagonal matrix with diagonal elements $\{x_1, \cdots, x_K\}$ and $\diag(\mathbf{X})$ returns a diagonal matrix having the main diagonal elements of $\mathbf{X}$ on its main diagonal. $\Re(\cdot)$ extracts the real part of a complex-valued input; the circularly symmetric complex Gaussian distribution with mean $\mu$ and variance $\sigma^2$ is denoted by ${\cal CN}(\mu,\sigma^2)$; and $\sim$ stands for ``distributed as".
\vspace*{-0mm}

\subsection{Cognitive Radio System Model}
\begin{figure}
\centering\vspace*{-0mm}
\includegraphics[width=2.8in]{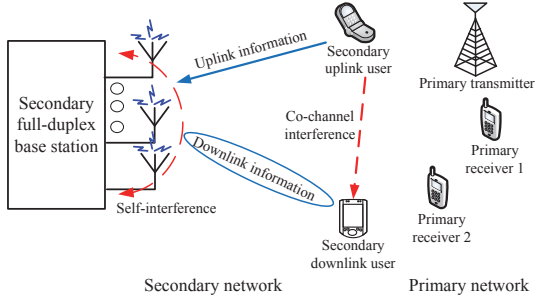}\vspace*{-0mm}
\caption{A CR system where a secondary FD BS, $K=1$ secondary HD downlink user, and $J=1$ secondary HD uplink user share the same spectrum with $R=2$ primary HD receivers.}
\label{fig:system_model}\vspace*{-4mm}
\end{figure}
 The considered CR system comprises one secondary FD BS, $K$ secondary DL users, $J$ secondary UL users, one primary transmitter, and $R$ primary receivers. The secondary FD BS is equipped with $N_{\mathrm{T}} > 1$ antennas for facilitating simultaneous DL transmission and UL reception in the secondary network in the same frequency band.
%The $K+J$ secondary users are single-antenna HD mobile communication devices to ensure low hardware complexity.
The $K+J$ secondary users, the primary transmitter, and the secondary receivers are single-antenna HD devices that share the same spectrum, cf. Figure 1.
The number of antennas at the secondary FD BS is assumed to be larger than the number of secondary UL users to facilitate reliable UL signal detection, i.e., $N_\mathrm{T} \ge J$. The secondary FD BS provides wireless service to the secondary users applying multiuser multiple-input multiple-output (MU-MIMO) techniques. The primary transmitter provides conventional broadcast services to the primary receivers.

% \subsection{Channel Model}
In this paper, we focus on slow frequency flat fading channels. In each scheduling time slot, the secondary FD BS transmits $K$ independent signal streams simultaneously at the same frequency to the $K$ secondary DL users. In particular, the information signal to secondary DL user $k \in \{1,\ldots,K\}$ can be expressed as $\mathbf{x}_k=\mathbf{w}_k d_k^{\mathrm{DL}}$, where $d_k^{\mathrm{DL}}\in\mathbb{C}$ and $\mathbf{w}_k\in\mathbb{C}^{N_\mathrm{T}\times1}$ are the information bearing signal for DL user $k$ and the corresponding beamforming vector, respectively. Without loss of generality, we assume ${\cal E}\{\abs{d_k^{\mathrm{DL}}}^2\}=1,\forall k\in\{1,\ldots,K\}$. Therefore, the received signal at secondary DL user $k\in\{1,\ldots,K\}$, the secondary FD BS, and primary receiver $r\in\{1,\ldots,R\}$ are given by \vspace*{-2mm}
%\begin{eqnarray}
%\label{eqn:dl_user_rcv_signal}y_{k}^{\mathrm{DL}}\hspace*{-2mm}&=&\hspace*{-2mm}\mathbf{h}_k^H\mathbf{x}_k\hspace*{3mm} +\hspace*{-0.5mm}\underbrace{\sum_{m\neq k}^K\mathbf{h}_k^H\mathbf{x}_m}_{\mbox{multiuser interference}}\hspace*{3mm} \notag\\
%&& + \hspace*{-0.5mm}\underbrace{\sum_{j=1}^J \sqrt{P_j}f_{j,k}d_j^{\mathrm{UL}}}_{\mbox{co-channel interference}}\hspace*{-0.5mm} + \hspace*{3mm} n^{\mathrm{DL}}_{k}\hspace*{-0.5mm},\,\,
%\end{eqnarray}\vspace*{-2mm}
\begin{eqnarray}
\label{eqn:dl_user_rcv_signal}y_{k}^{\mathrm{DL}}\hspace*{-0mm}=\hspace*{-0mm}\mathbf{h}_k^H\mathbf{x}_k\hspace*{0mm} +\hspace*{-0mm}\underbrace{\sum_{m\neq k}^K\mathbf{h}_k^H\mathbf{x}_m}_{{\underset{\mbox{interference}} {\mbox{multiuser}}}}\hspace*{0mm}
+ \hspace*{-0mm}\underbrace{\sum_{j=1}^J \sqrt{P_j}f_{j,k}d_j^{\mathrm{UL}}}_{{\underset{\mbox{interference}} {\mbox{co-channel}}}}\hspace*{0.5mm} + \hspace*{0.5mm} n^{\mathrm{DL}}_{k}\hspace*{-0mm},\,\,
\end{eqnarray}\vspace*{-5mm}
%%%%%%%%
\begin{eqnarray}
\hspace*{-12.5mm}\label{eqn:ul_rcv_signal}\mathbf{y}^{\mathrm{UL}}\hspace*{-2mm}&=&\hspace*{-2mm}\sum_{j=1}^J \sqrt{P_j}\mathbf{g}_j d_j^{\mathrm{UL}}\hspace*{1mm} +\hspace*{-2mm}\underbrace{\mathbf{H}_{\mathrm{SI}}\sum_{k=1}^K{\mathbf{x}_k}}_{\mbox{self-interference}}\hspace*{-2mm} +\hspace*{1mm}\mathbf{n}^{\mathrm{UL}} ,\,\, \text{and}
\end{eqnarray}\vspace*{-5mm}
%%%%%%%%
\begin{eqnarray}
\hspace*{-15mm}\label{eqn:primary_rcv_signal}y^{\mathrm{PU}}_r\hspace*{-2mm}&=& \hspace*{-0.5mm}\sum_{k=1}^K\mathbf{l}_r^H\mathbf{x}_k \hspace*{1mm} +\hspace*{1mm}\sum_{j=1}^J \sqrt{P_j}e_{j,r} d_j^{\mathrm{UL}}\hspace*{2mm} +\hspace*{2mm}n^{\mathrm{PU}}_r\hspace*{0mm},\,\,
\end{eqnarray}
respectively. The DL channel between the secondary FD BS and secondary DL user $k$ is denoted by $\mathbf{h}_k\in\mathbb{C}^{N_{\mathrm{T}}\times1}$ and $f_{j,k}\in\mathbb{C}$ represents the channel between secondary UL user $j$ and secondary DL user $k$. Variables $d_j^{\mathrm{UL}}$, ${\cal E}\{\abs{d_j^{\mathrm{UL}}}^2\}=1$, and $P_j$ are the data and transmit power sent from secondary UL user $j$ to the secondary FD BS, respectively. Vector $\mathbf{g}_j\in\mathbb{C}^{N_{\mathrm{T}}\times1}$ denotes the channel between secondary UL user $j$ and the secondary FD BS. Matrix $\mathbf{H}_{\mathrm{SI}}\in{\mathbb{C}^{N_{\mathrm{T}}\times N_{\mathrm{T}}}}$ denotes the self-interference (SI) channel of the secondary FD BS. The SI is caused by the signal leakage from DL transmission to UL reception in the secondary network. Vector $\mathbf{l}_r \in{\mathbb{C}^{N_{\mathrm{T}}\times 1}}$ denotes the channel between the secondary FD BS and primary receiver $r$. Scalar $e_{j,r}\in\mathbb{C}$ denotes the channel between secondary UL user $j$ and primary receiver $r$. Variables $\mathbf{h}_k$, $f_{j,k}$, $\mathbf{g}_j$, $\mathbf{H}_{\mathrm{SI}}$, $\mathbf{l}_r$, and $e_{j,r}$ capture the joint effect of path loss and small scale fading.  $\mathbf{n}^{\mathrm{UL}}\sim{\cal CN}(\mathbf{0},\sigma_{\mathrm{UL}}^2\mathbf{I}_{N_\mathrm{T}})$ and $n^{\mathrm{DL}}_{k}\sim{\cal CN}(0,\sigma_{\mathrm{n}_k}^2)$ are the equivalent noises at the secondary FD BS and secondary DL user $k$, which capture the joint effect of the received interference from the primary transmitter and thermal noise. $n^{\mathrm{PU}}_{r}\sim{\cal CN}(0,{\sigma^2_{\mathrm{PU}_r}})$ represent the additive white Gaussian noise (AWGN) at primary receiver $r$.
In (\ref{eqn:dl_user_rcv_signal}), the term $\sum_{j=1}^J \sqrt{P_j}f_{j,k}d_j^{\mathrm{UL}}$ denotes the aggregated co-channel interference (CCI) caused by the UL users to DL user $k$. In (\ref{eqn:ul_rcv_signal}), the term $\mathbf{H}_{\mathrm{SI}}\sum_{k=1}^K\mathbf{x}_k$ represents the SI.

\vspace*{0mm}

\section{Resource Allocation Problem Formulation}
In this section, we formulate the resource allocation design as a non-convex optimization problem, after introducing the adopted performance metrics and the CSI assumed for resource allocation. For the sake of notational simplicity, we define the following variables: $\mathbf{H}_k=\mathbf{h}_k\mathbf{h}_k^H$, $k\in\{1,\ldots,K\}$, $\mathbf{G}_j=\mathbf{g}_j\mathbf{g}_j^H$, $j\in\{1,\ldots,J\}$, and $\mathbf{V}_j=\mathbf{v}_j\mathbf{v}_j^H$, $j\in\{1,\ldots,J\}$.
\vspace*{-2mm}
\subsection{Performance Metrics}
\vspace*{-0mm}
The receive signal-to-interference-plus-noise ratio (SINR) at secondary DL user $k$ is given by \vspace*{-2mm}
\begin{eqnarray}
\Gamma^{\mathrm{DL}}_{k}=\frac{\abs{\mathbf{h}_k^H\mathbf{w}_k}^2}{\overset{K}{\underset{m \neq k}{\sum}}\abs{\mathbf{h}_k^H\mathbf{w}_m}^2 + \overset{J}{\underset{j=1}{\sum}}P_j\abs{f_{j,k}}^2 + \sigma_{\mathrm{n}_k}^2}.
\end{eqnarray}
On the other hand, the receive SINR of secondary UL user $j$ at the secondary FD BS is given by \vspace*{-2mm}
%\begin{eqnarray}
%\Gamma^{\mathrm{UL}}_{j}=\frac{P_j\abs{\mathbf{g}_j^H\mathbf{v}_j}^2}{\overset{J}{\underset{n \neq j}{\sum}}P_n\abs{\mathbf{g}_n^H\mathbf{v}_j}^2+\overset{K}{\underset{k=1}{\sum}}\abs{\mathbf{v}_j^H\mathbf{H}_{\mathrm{SI}}\mathbf{w}_k}^2 +\sigma_{\mathrm{UL}}^2\norm{\mathbf{v}_j}^2},
%\end{eqnarray}
\begin{eqnarray}
\Gamma^{\mathrm{UL}}_{j}=\frac{P_j\abs{\mathbf{g}_j^H\mathbf{v}_j}^2}{\overset{J}{\underset{n \neq j}{\sum}}\hspace*{-1mm}P_n\abs{\mathbf{g}_n^H\mathbf{v}_j}^2\hspace*{-1mm} + I_j^{\mathrm{SI}} +\hspace*{-1mm}\sigma_{\mathrm{UL}}^2\norm{\mathbf{v}_j}^2},
\end{eqnarray}
where $\mathbf{v}_j\in\mathbb{C}^{N_{\mathrm{T}}\times1}$ is the receive beamforming vector for decoding the information received from  secondary UL user $j$. Besides, we define $I_j^{\mathrm{SI}}=\hspace*{-1mm}\Tr\hspace*{-1mm}\big(\hspace*{-0.5mm}\rho\mathbf{V}_j \hspace*{-0.5mm}\diag\hspace*{-1mm}\big(\hspace*{-0.5mm}{\sum_{k=1}^{K}}\hspace*{-0.5mm} \mathbf{H}_{\mathrm{SI}} \mathbf{w}_k\hspace*{-0.5mm}\mathbf{w}^H_k\hspace*{-0.5mm}\mathbf{H}_{\mathrm{SI}}^H\big)\big)$, where $0 <\rho \ll 1$ is a constant modelling the noisiness of the SI cancellation at the secondary FD BS {\cite[Eq. (4)]{JR:FD_model}}. In this paper, we adopt zero-forcing receive beamforming (ZF-BF) \cite{book:david_wirelss_com} as it approaches the performance of optimal minimum mean square error beamforming (MMSE-BF) when the noise term is not dominating \cite{book:david_wirelss_com} or the number of antennas is sufficiently large \cite{Mpair_FDRelay_massive}. Besides, ZF-BF facilitates the design of a computational efficient resource allocation algorithm.
\vspace*{-1mm}
\subsection{Channel State Information}
In this paper, we assume that the CSI of all secondary users is perfectly known at the secondary BS because of frequent channel estimation. However, for the secondary network-to-primary network channels, the perfect CSI assumption may not hold since the primary receivers do not interact directly with the secondary network. Hence, the CSI of the link between the secondary FD BS and primary receiver $r\in\{1,\ldots,R\}$, i.e., $\mathbf{l}_{r}$, and the link between the secondary UL user $j\in\{1,\ldots,J\}$ and primary receiver $r$, i.e., $e_{j,r}$, are modeled as \vspace*{-2mm}
\begin{eqnarray}
%%%n%
\hspace*{-2mm} \label{eqn:BS-pri-channel}\mathbf{l}_{r}\hspace*{-2mm}&=&\hspace*{-2mm}\mathbf{\hat l}_{r} + \Delta\mathbf{l}_{r},\,
\label{eqn:BS-pri-set}\mathbf{\Omega}_{\mathrm{DL}_{r}} \hspace*{-1mm}\triangleq \hspace*{-0.5mm} \Big\{\mathbf{l}_{r}\in \mathbb{C}^{N_{\mathrm{T}}\times 1} \hspace*{-0.5mm} : \hspace*{-0.5mm}\norm{\Delta\mathbf{l}_{r}} \hspace*{-1mm} \le \hspace*{-0.5mm} \varepsilon_{\mathrm{DL}_{r}}\Big\},\\[-2mm]
%%%%%
\label{eqn:UL-pri-channel}
\hspace*{-2mm} e_{j,r}\hspace*{-2mm}&=&\hspace*{-2mm}{\hat e}_{j,r}\hspace*{-1mm}  + \hspace*{-1mm} \Delta e_{j,r},\,
\label{eqn:UL-pri-set}
{\Omega }_{\mathrm{UL}_{j,r}} \hspace*{-1mm}\triangleq \hspace*{-0.5mm}\Big\{ e_{j,r}\in \mathbb{C} \hspace*{-0.5mm} : \hspace*{-0.5mm}\abs{\Delta e_{j,r}}\hspace*{-1mm} \le \hspace*{-0.5mm} \varepsilon_{\mathrm{UL}_{j,r}}\Big\},
\end{eqnarray}
respectively, where ${\hat e}_{j,r}$ and $\mathbf{\hat l}_{r}$ are the CSI estimates, and $\Delta e_{j,r}$ and $\Delta\mathbf{l}_{r}$ denote the unknown CSI estimation errors. The continuous sets ${\Omega }_{\mathrm{UL}_{j,r}}$ and $\mathbf{\Omega}_{\mathrm{DL}_{r}}$ contain all possible channel uncertainties, and $\varepsilon_{\mathrm{UL}_{j,r}}$ and $\varepsilon_{\mathrm{DL}_{r}}$ denote the bounded magnitude of ${\Omega }_{\mathrm{UL}_{j,r}}$ and $\mathbf{\Omega}_{\mathrm{DL}_{r}}$, respectively.
%In practice, the values of $\varepsilon_{\mathrm{UL}_{j,r}}$ and $\varepsilon_{\mathrm{DL}_{r}}$ depend on the coherence time of the associated channels and the transmission duration of the scheduling slot.
\vspace*{-3mm}
\subsection{Optimization Problem Formulation}
The system objective is to minimize the maximum interference leakage from the secondary network to the primary receivers. The optimal power allocation and beamformer design are obtained by solving the following optimization problem: \vspace*{-3mm}
%\begin{eqnarray}
%\label{eqn:pro}
%&&\hspace*{-10mm} \underset{\mathbf{w}_k,P_j}{\mino}\,\,\hspace*{-2mm} \,\,\underset{r\in\{1,\ldots,R\}}{\underset{\Delta e_{j,r} \in \mathbf{\Omega}_{\mathrm{UL}_{j,r}},\Delta \mathbf{l}_r\in \mathbf{\Omega}_{\mathrm{DL}_r},}{\max}}\,\,\hspace*{-2mm}\,\, \sum_{k=1}^{K}\norm{\mathbf{l}_r^H\mathbf{w}_k}^2+\sum_{j=1}^{J}P_j\abs{e_{j,r}}^2 \notag \\[-0mm]
%\notag\mbox{s.t.}
%&&\hspace*{-5mm}\mbox{C1: } \frac{\abs{\mathbf{h}_k^H\mathbf{w}_k}^2}{\overset{K}{\underset{m \neq k}{\sum}}\abs{\mathbf{h}_k^H\mathbf{w}_m}^2 + \overset{J}{\underset{j=1}{\sum}}P_j\abs{f_{j,k}}^2 +\sigma_{\mathrm{n}_k}^2} \geq \Gamma^{\mathrm{DL}}_{\mathrm{req}_k},\,\, \forall k, \notag\\[-0mm]
%%%%%
%&&\hspace*{-5mm}\mbox{C2: }\frac{P_j\abs{\mathbf{g}_j^H\mathbf{v}_j}^2}{\overset{J}{\underset{n \neq j}{\sum}}\hspace*{-1mm}P_n\abs{\mathbf{g}_n^H\mathbf{v}_j}^2\hspace*{-1mm} + I_j^{\mathrm{SI}} +\hspace*{-1mm}\sigma_{\mathrm{UL}}^2\norm{\mathbf{v}_j}^2} \hspace*{-1mm}\geq\hspace*{-1mm} \Gamma^{\mathrm{UL}}_{\mathrm{req}_j},\,\,\hspace*{-1.5mm} \forall j,  \notag\\
%%%%
%&&\hspace*{-5mm}\mbox{C3: }\sum_{k=1}^{K}\norm{\mathbf{w}_k}^2 \le P^{\mathrm{max}}_{\mathrm{DL}}, \quad \mbox{C4: } 0 \le P_j \le P^{\mathrm{max}}_{\mathrm{UL}_j},\,\, \forall j.
%\end{eqnarray}
\begin{eqnarray}
\label{eqn:pro}
&&\hspace*{-10mm} \underset{\mathbf{w}_k,P_j}{\mino}\,\,\hspace*{-2mm} \,\,\underset{r\in\{1,\ldots,R\}}{\underset{\Delta e_{j,r} \in \mathbf{\Omega}_{\mathrm{UL}_{j,r}},\Delta \mathbf{l}_r\in \mathbf{\Omega}_{\mathrm{DL}_r},}{\max}}\,\,\hspace*{-2mm}\,\, \sum_{k=1}^{K}\norm{\mathbf{l}_r^H\mathbf{w}_k}^2+\sum_{j=1}^{J}P_j\abs{e_{j,r}}^2 \notag \\[-1mm]
\notag\mbox{s.t.}
&&\hspace*{-5mm}\mbox{C1: } \Gamma^{\mathrm{DL}}_{k} \geq \Gamma^{\mathrm{DL}}_{\mathrm{req}_k},\,\, \forall k,  \quad
\hspace*{3mm}\mbox{C2: }\Gamma^{\mathrm{UL}}_{j} \hspace*{-1mm}\geq\hspace*{-1mm} \Gamma^{\mathrm{UL}}_{\mathrm{req}_j},\,\,\hspace*{-0mm} \forall j,  \notag\\[-2mm]
%%%
&&\hspace*{-5mm}\mbox{C3: }\sum_{k=1}^{K}\norm{\mathbf{w}_k}^2 \le P^{\mathrm{max}}_{\mathrm{DL}}, \quad \mbox{C4: } 0 \le P_j \le P^{\mathrm{max}}_{\mathrm{UL}_j},\,\, \forall j.
\end{eqnarray}
Constants $\Gamma^{\mathrm{DL}}_{\mathrm{req}_k} > 0$ and $\Gamma^{\mathrm{UL}}_{\mathrm{req}_j} > 0$ in constraints C1 and C2 in (\ref{eqn:pro}) are the minimum required SINR for secondary DL users $k\in\{1,\ldots,K\}$ and secondary UL users $j\in\{1,\ldots,J\}$, respectively.
Constants $P^{\mathrm{max}}_{\mathrm{DL}}> 0$ and $P^{\mathrm{max}}_{\mathrm{UL}_j}> 0$ in constraints C3 and C4 in (\ref{eqn:pro}) are the maximum transmit power allowance for the secondary FD BS and secondary UL users $j\in\{1,\ldots,J\}$, respectively.
The problem in (\ref{eqn:pro}) is a non-convex problem due to the non-convex constraints C1 and C2. Besides, the objective function of (\ref{eqn:pro}) involves infinitely many functions due to the continuity of the CSI uncertainty sets.

\vspace*{-1mm}
\section{Solution of the Optimization Problem}
To solve the non-convex problem in (\ref{eqn:pro}) efficiently, we first reformulate the problem in an equivalent form and then transform the non-convex constraints into equivalent linear matrix inequality (LMI) constraints. Finally, the problem is solved by semidefinite programming (SDP) relaxation.

To facilitate the SDP relaxation, we define $\mathbf{W}_k=\mathbf{w}_k\mathbf{w}_k^H$ and rewrite the problem in the following equivalent form:\vspace*{-3mm}
\begin{eqnarray}
\label{eqn:SDP-pro}
&&\hspace*{-1mm} \underset{\mathbf{W}_k\in \mathbb{H}^{N_\mathrm{T}},P_j,\tau}{\mino}\,\,\hspace*{0mm} \,\, \tau \notag \\[-1mm]
\notag\mbox{s.t.}
&&\hspace*{-5mm}\mbox{C1: } \frac{\Tr(\mathbf{H}_k\mathbf{W}_k)}{\Gamma^{\mathrm{DL}}_{\mathrm{req}_k}} \ge {I_k^{\mathrm{DL}}+\sigma_{{\mathrm{n}}_k}^2},\,\, \forall k,\\[-1mm]
%%%%
&&\hspace*{-5mm}\mbox{C2: }\frac{P_j\Tr(\mathbf{V}_j\mathbf{G}_j)}{\Gamma^{\mathrm{UL}}_{\mathrm{req}_j}}
 \ge {I_j^\mathrm{UL}+\sigma_{\mathrm{UL}}^2\Tr(\mathbf{V}_j)},\,\, \forall j,\notag\\[-3mm]
%%%
&&\hspace*{-5mm}\mbox{C3: }\sum_{k=1}^{K}\Tr(\mathbf{W}_k) \le P^{\mathrm{max}}_{\mathrm{DL}}, \quad \mbox{C4: } 0 \le P_j \le P^{\mathrm{max}}_{\mathrm{UL}_j},\,\, \forall j,\notag\\[-4mm]
%%%
&&\hspace*{-5mm}\mbox{C5: } \underset{r\in\{1,\ldots,R\}}{\underset{\Delta \mathbf{l}_r\in \mathbf{\Omega}_{\mathrm{DL}_r},}{\underset{\Delta e_{j,r} \in \Omega_{\mathrm{UL}_{j,r}},}{\max}}}\,\,\hspace*{-2mm}\,\, \sum_{k=1}^{K}\mathbf{l}_r^H \mathbf{W}_k \mathbf{l}_r+\sum_{j=1}^{J}P_j\abs{e_{j,r}}^2 \le \tau ,\,\, \notag\\
%%%
&&\hspace*{-5mm}\mbox{C6: }\mathbf{W}_k \succeq \mathbf{0} ,\,\, \forall k, \quad \mbox{C7: }\Rank(\mathbf{W}_k) \le 1 ,\,\, \forall k,
\end{eqnarray}
where $\mathbf{W}_k\succeq \mathbf{0}$, ${\mathbf{W}_k}\in \mathbb{H}^{N_\mathrm{T}}$, and $\Rank(\mathbf{W}_k) \le 1$ in (\ref{eqn:SDP-pro}) are imposed to guarantee that $\mathbf{W}_k=\mathbf{w}_k\mathbf{w}_k^H$ holds after optimization.
Furthermore, we use $I_k^\mathrm{DL}\hspace*{-1mm}=\hspace*{-1mm} \sum_{m \neq k}^{K}\Tr(\mathbf{H}_k\mathbf{W}_m) + \sum_{j=1}^{J}P_j\abs{f_{j,k}}^2$ and
$I_j^\mathrm{UL}\hspace*{-1mm}=\hspace*{-1mm}\Tr\hspace*{-1mm}\big(\hspace*{-0.5mm}\rho\mathbf{V}_j \hspace*{-0.5mm}\diag\hspace*{-1mm}\big(\hspace*{-0.5mm}{\sum_{k=1}^{K}}\hspace*{-0.5mm} \mathbf{H}_{\mathrm{SI}} \mathbf{W}_k\hspace*{-0.5mm}\mathbf{H}_{\mathrm{SI}}^H\big)\big) + \hspace*{-1mm}\sum_{n \neq j}^{J}\hspace*{-0.5mm}P_n\Tr(\mathbf{G}_n\hspace*{-0.5mm}\mathbf{V}_j)\hspace*{-0mm}.$
%\begin{eqnarray}
%I_k^\mathrm{DL}\hspace*{-2mm}&=&\hspace*{-2mm} \overset{K}{\underset{m \neq k}{\sum}}\Tr(\mathbf{H}_k\mathbf{W}_m) + \overset{J}{\underset{j=1}{\sum}}P_j\abs{f_{j,k}}^2,\,\, \text{and}\\[-1mm]
%I_j^\mathrm{UL}\hspace*{-2mm}&=&\hspace*{-2mm}\overset{J}{\underset{n \neq j}{\sum}}\hspace*{-0.5mm}P_n\Tr(\mathbf{G}_n\hspace*{-0.5mm}\mathbf{V}_j)\hspace*{-1mm}+\hspace*{-1.5mm}\overset{K}{\underset{k=1}{\sum}} \hspace*{-0.5mm}\Tr(\mathbf{V}_j\hspace*{-0.5mm}\mathbf{H}_{\mathrm{SI}}\mathbf{W}_k\mathbf{H}_{\mathrm{SI}}^H).
%\end{eqnarray}
$\tau$ is an auxiliary optimization variable and (\ref{eqn:SDP-pro}) is the epigraph representation of (\ref{eqn:pro}).
Constraint ${{\mbox{C5}}}$ involves an infinite number of inequality constraints, as the estimation error variables $\Delta{e}_{j,r}$ and $\Delta\mathbf{l}_r$ are involved. Here, we introduce a scalar slack variable $\delta_r$ to handle the coupled estimation error variables in constraint ${\mbox{C5}}$. In particular, constraint ${\mbox{C5}}$ can be equivalently represented by \vspace*{-2mm}
\begin{eqnarray}
&&\hspace*{-5mm}{{\mbox{C5a}}}\mbox{: } \sum_{k=1}^{K}\mathbf{l}_r^H \mathbf{W}_k \mathbf{l}_r \le \delta_r,\,\, \forall \mathbf{l}_r \in \mathbf{\Omega}_{\mathrm{DL}_r},\, \forall r,\\[-3mm]
&&\hspace*{-5mm}{{\mbox{C5b}}}\mbox{: }\delta_r \le \tau - \sum_{j=1}^{J}P_j\abs{e_{j,r}}^2,\,\,  \forall e_{j,r} \in \mathbf{\Omega}_{\mathrm{UL}_{j,r}},\, \forall j, r.
\end{eqnarray}

Now, we introduce a lemma which allows us to transform constraint ${\mbox{C5a}}$ into an LMI.
\begin{Lem}[S-Procedure \cite{book:convex}] Let a function $f_m(\mathbf{x}),m\in\{1,2\},\mathbf{x}\in \mathbb{C}^{N\times 1},$ be defined as \vspace*{-2mm}
\begin{eqnarray}
\label{eqn:S-proc-function}f_m(\mathbf{x})=\mathbf{x}^H\mathbf{A}_m\mathbf{x}+2 \hspace*{0mm}\Re\hspace*{0mm} \{\mathbf{b}_m^H\mathbf{x}\}+c_m,
\end{eqnarray}
where $\mathbf{A}_m\in\mathbb{H}^N$, $\mathbf{b}_m\in\mathbb{C}^{N\times 1}$, and $c_m\in\mathbb{R}^{1\times 1}$. Then, the implication $f_1(\mathbf{x})\le 0\Rightarrow f_2(\mathbf{x})\le 0$  holds if and only if there exists a variable $\delta\ge 0$ such that \vspace*{-2mm}
\begin{eqnarray}\label{eqn:S-proc-LMI}\delta
\begin{bmatrix}
       \mathbf{A}_1 & \mathbf{b}_1          \\
       \mathbf{b}_1^H & c_1           \\
           \end{bmatrix} -\begin{bmatrix}
       \mathbf{A}_2 & \mathbf{b}_2          \\
       \mathbf{b}_2^H & c_2           \\
           \end{bmatrix}          \succeq \mathbf{0}         ,
\end{eqnarray}
provided that there exists a point $\mathbf{\hat{x}}$ such that $f_k(\mathbf{\hat{x}})<0$.
\end{Lem}

By applying (\ref{eqn:BS-pri-channel}), constraint ${\mbox{C5a}}$ can be equivalently expressed as \vspace*{-5mm}
\begin{eqnarray}
{\mbox{C5a}}\mbox{: }
0 \hspace*{-2mm} &\ge& \hspace*{-2mm} \Delta\mathbf{l}^H_{r} \overset{K}{\underset{k=1}{\sum}}\mathbf{W}_k \Delta\mathbf{l}_{r}\hspace*{-0.5mm} \notag \\[-2mm]
&+& \hspace*{-2mm}2\Re\{\mathbf{\hat  l}^H_r \overset{K}{\underset{k=1}{\sum}}\mathbf{W}_k \Delta\mathbf{l}_r\}
\hspace*{0mm} + \hspace*{0mm}\mathbf{\hat l}^H_r\overset{K}{\underset{k=1}{\sum}}\mathbf{W}_k\mathbf{\hat l}_r\hspace*{-0mm} - \hspace*{-0mm} \delta_r
.\,\, \notag
\end{eqnarray}
By exploiting Lemma 1, we obtain the following implications: \vspace*{-0mm}
$\Delta \mathbf{l}_{r}^H\Delta \mathbf{l}_{r} \hspace*{-0.5mm} - \hspace*{-0.5mm} \varepsilon_{\mathrm{DL}_{r}}^2 \hspace*{-0.5mm} \le \hspace*{-0.5mm} 0
\Rightarrow   {\mbox{C5a}}$
%\begin{eqnarray}
%\label{eqn:CCI-err-QC}\Delta \mathbf{l}_{r}^H\Delta \mathbf{l}_{r} \hspace*{-0.5mm} - \hspace*{-0.5mm} \varepsilon_{\mathrm{DL}_{r}}^2 \hspace*{-0.5mm} \le \hspace*{-0.5mm} 0
%\Rightarrow   {\mbox{C5a}} \notag
%\end{eqnarray}
holds if and only if there exists a variable $\alpha_r \ge 0$ such that \vspace*{-2mm}
\begin{eqnarray}\label{eqn:LMI-C5a}
\overline{\text{C5}}\mbox{a: }\mathbf{R}_{\overline{\mathrm{C5}}\mathrm{a}_r}\big(\mathbf{W}_k,\alpha_r,\delta_r \big)\hspace*{40mm}\notag\\[-2mm]
=
          \begin{bmatrix}
       \alpha_r\mathbf{I}_{N_{\mathrm{T}}}\hspace*{-0.5mm}  &     \hspace*{-1mm} \mathbf{0}          \\
       \mathbf{0}    \hspace*{-1mm}                         &     \hspace*{-1mm} -\alpha_r\varepsilon_{\mathrm{DL}_r}^2 + \delta_r\\
           \end{bmatrix}
             \hspace*{-0.5mm}-\hspace*{-0.5mm} \mathbf{B}_{\mathbf{l}_r}^H\overset{K}{\underset{k=1}{\sum}}\mathbf{W}_k\hspace*{-0.5mm}\mathbf{B}_{\mathbf{l}_r} \succeq \mathbf{0}, \forall k, r,
\end{eqnarray}
holds, where $\mathbf{B}_{\mathbf{l}_r}=\big[\mathbf{I}_{N_{\mathrm{T}}}\quad \hat{\mathbf{l}}_r\big]$.
Similarly, by applying Lemma 1 to constraint C5b, we obtain an equivalent constraint \vspace*{-2mm}
\begin{eqnarray}\label{eqn:LMI-C5a}
\overline{\text{C5}}\mbox{b: }\mathbf{R}_{\overline{\mathrm{C5}}b_r}\big(P_j,\beta_r,\delta_r,\tau \big)\hspace*{38mm}\notag\\[-1mm]
=
          \begin{bmatrix}
       \beta_r\mathbf{I}_{N_{\mathrm{T}}}-\mathbf{P}\hspace*{-0.5mm}  &     \hspace*{-1mm} -\mathbf{P}{\mathbf{\hat e}}_r          \\
       -{\mathbf{\hat e}}_r^H\mathbf{P}    \hspace*{-1mm}             &     \hspace*{-1mm} -\beta_r\varepsilon_{\mathrm{UL}_r}^2 - \delta_r + \tau -{\mathbf{\hat e}}_r^H\mathbf{P}{\mathbf{\hat e}}_r\\
           \end{bmatrix}
              \succeq \mathbf{0}, \forall r,
\end{eqnarray}
where $\beta_r \ge 0$, $\mathbf{P}\hspace*{-0.5mm}=\hspace*{-0.5mm}\diag\hspace*{-0.5mm}\big(P_1,\ldots,P_J\big)$, and $\mathbf{\hat e}_r\hspace*{-0.5mm}=\hspace*{-0.5mm}\big[\hat e_{1,r},\ldots,\hat e_{J,r}\big]^T\hspace*{-1mm}$.

Next, we relax the non-convex constraint C7: $\Rank(\mathbf{W}_k) \le 1$ by removing it from the problem formulation such that the considered problem becomes a convex SDP: \vspace*{-3mm}
\begin{eqnarray}
\label{eqn:SDR-robust}&&\hspace*{0mm} {\underset{\mathbf{W}_k\in\mathbb{H}^{N_\mathrm{T}},P_j,\tau,\delta_r, \alpha_{r},\beta_{r}}{\mino}}\,\,\,\, \tau \notag \\[-1mm]
%%% C1, C2, C3, C4, C6
\notag \mbox{s.t.} &&\hspace*{-5mm}{\mbox{C1}},{\mbox{C2}},{\mbox{C3}},{\mbox{C4}},{\mbox{C6}}, \quad \hspace*{0mm}\mbox{C8: }\delta_r,\alpha_{r}, \beta_{r} \ge 0,\forall r,\\[-1mm]
%%% C5a
&&\hspace*{-5mm}\overline{\text{C5}}\mbox{a: }\mathbf{R}_{\overline{\mathrm{C5}}a_r}\big(\mathbf{W}_k,\alpha_r,\delta_r \big)\succeq \mathbf{0},\forall r, \notag  \\[-1mm]
%%%C5b
&& \hspace*{-5mm}\overline{\text{C5}}\mbox{b: }\mathbf{R}_{\overline{\mathrm{C5}}b_r}\big(P_j,\beta_r,\delta_r,\tau \big)\succeq \mathbf{0},\forall r.
%%% C8
%&& \hspace*{-5mm}\mbox{C8: }\delta_r,\alpha_{r}, \beta_{r} \ge 0,\forall r.
\end{eqnarray}
The relaxed convex problem in (\ref{eqn:SDR-robust}) can be solved efficiently by standard convex program solvers such as CVX \cite{website:CVX}. Besides, if the solution obtained for a relaxed SDP problem is a rank-one matrix, i.e., $\Rank(\mathbf{W}_k) = 1$ for $\mathbf{W}_k\ne\mathbf{0}, \, \forall k$, then it is also the optimal solution of the original problem. Next, we reveal the tightness of the SDP relaxation in the following theorem.
\begin{Thm}\label{thm:rankone_condition}
Assuming the considered problem is feasible, for $\Gamma^{\mathrm{DL}}_{\mathrm{req}_k} > 0$, we can always obtain or construct an optimal rank-one matrix $\mathbf{W}_k^*$.
\end{Thm}
\emph{\quad Proof: } Please refer to the Appendix. \hfill\qed

\vspace*{-0mm}
\section{Results}

\begin{table}[t]\caption{System parameters.}\label{tab:parameters} \vspace*{-2mm}
\newcommand{\tabincell}[2]{\begin{tabular}{@{}#1@{}}#2\end{tabular}}
\centering
\begin{tabular}{|l|l|}\hline
\hspace*{-1mm}Carrier center frequency & \mbox{$1.9$ GHz}  \\
\hline
\hspace*{-1mm}System bandwidth & \mbox{$200$ kHz}  \\
\hline
\hspace*{-1mm}Path loss exponent &  \mbox{$3.6$}  \\
\hline
%\hspace*{-1mm}Cell radius &  \mbox{$250$ m}  \\
%\hline
\hspace*{-1mm}SI cancellation    &  \mbox{$-80$ dB}    \\
\hline
\hspace*{-1mm}Secondary DL user equivalent noise power, $\sigma_{\mathrm{n}_k}^2$ &  \mbox{$-90$ dBm}   \\
\hline
\hspace*{-1mm}Secondary FD BS equivalent noise power, $\sigma_{\mathrm{UL}}^2$ &  \mbox{$-90$ dBm}  \\
\hline
\hspace*{-1mm}Secondary FD BS antenna gain &  \mbox{$10$ dBi}  \\
\hline
\hspace*{-1mm}Max. transmit power at the secondary FD BS, $P^{\mathrm{max}}_{\mathrm{DL}}$ &  \mbox{$30$ dBm}  \\
\hline
\hspace*{-1mm}Max. transmit power at the secondary UL users, $P^{\mathrm{max}}_{\mathrm{UL}}$ &  \mbox{$10$ dBm}  \\
\hline
\hspace*{-1mm}Max. transmit power at the primary transmitter &  \mbox{$30$ dBm}  \\
\hline
\end{tabular} \vspace*{-2mm}
\end{table}
In this section, we investigate the performance of the proposed resource allocation scheme through simulations. The most important simulation parameters are specified in Table \ref{tab:parameters}. There are $K=3$ secondary DL users, $J=5$ secondary UL users, and $R=2$ primary receivers in the system. We assume that the primary transmitter is $100$ meters away from the secondary FD BS. The secondary users and primary receivers are randomly and uniformly distributed between the reference distance of $5$ meters and the maximum service distance of $50$ meters of the corresponding secondary FD BS and primary transmitter, respectively.
%The FD BS is equipped with $N_\mathrm{T}$ antennas.
The small scale fading of the secondary DL channels, secondary UL channels, CCI channels, and secondary network-to-primary network channels are modeled as independent and identically distributed Rayleigh fading. The multipath fading coefficients of the SI channel are generated as independent and identically distributed Rician random variables with Rician factor $5$ dB. To facilitate the presentation, we define the maximum normalized estimation error of the secondary FD BS-to-primary receiver channel and the secondary UL user-to-primary receiver channel as $\frac{\varepsilon_{\mathrm{DL}_{r}}^2}{\norm{\mathbf{l}_{r}}^2}=\kappa_{\mathrm{DL}_{r}}^2$ and
$\frac{\varepsilon_{\mathrm{UL}_{j,r}}^2}{\norm{e_{j,r}}^2}=\kappa_{\mathrm{UL}_{j,r}}^2$, respectively. Besides, we assume that all channels have the same maximum normalized estimation error, i.e., $\kappa_{\mathrm{DL}_{m}}^2=\kappa_{\mathrm{UL}_{j,m}}^2=\kappa_{\mathrm{est}}^2$.
Furthermore, we assume that all secondary DL users and all secondary UL users require the same minimum SINRs, respectively, i.e., $\Gamma^{\mathrm{DL}}_{\mathrm{req}_k}=\Gamma^{\mathrm{DL}}_{\mathrm{req}}$ and $\Gamma^{\mathrm{UL}}_{\mathrm{req}_j}=\Gamma^{\mathrm{UL}}_{\mathrm{req}}$.

%In addition, we assume that all DL users and all UL users require the same minimum SINRs, respectively, i.e., $\Gamma^{\mathrm{DL}}_{\mathrm{req}_k}=\Gamma^{\mathrm{DL}}_{\mathrm{req}}$ and $\Gamma^{\mathrm{UL}}_{\mathrm{req}_j}=\Gamma^{\mathrm{UL}}_{\mathrm{req}}$.

\vspace*{-0mm}
% \subsection{Average Maximum Interference Leakage versus Minimum Required SINR}
In Figure 2, we investigate the average maximum interference leakage versus the minimum required secondary DL SINR, $\Gamma^{\mathrm{DL}}_{\mathrm{req}}$, for a minimum required secondary UL SINR of $\Gamma^{\mathrm{UL}}_{\mathrm{req}}=6$ dB, a maximum normalized channel estimation error of $\kappa_{\mathrm{est}}^2=5\%$, and different numbers of antennas at the secondary FD BS.
%%%%%%%%%
It can be observed that the average maximum interference leakage caused by the secondary network depends only weakly on  $\Gamma^{\mathrm{DL}}_{\mathrm{req}}$ due to the proposed robust optimization. Besides, Figure 2 also indicates that the interference leakage can be significantly reduced by increasing the number of secondary BS antennas. This is due to the fact that the extra degrees of freedom (DoF) offered by the additional antennas facilitate a more accurate DL beam-steering.
\begin{figure}
\centering\vspace*{-0mm}
\includegraphics[width=3.4in]{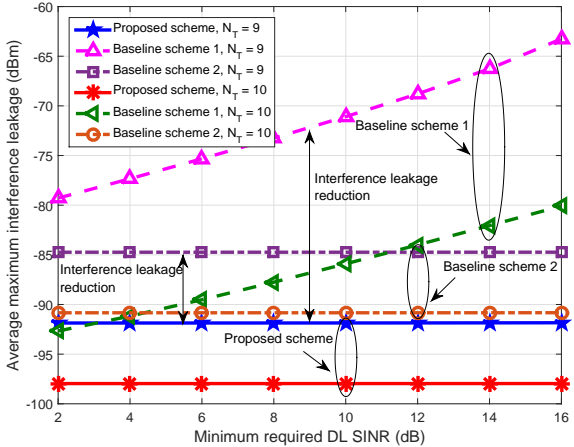}\vspace*{-3mm}
\caption{Average maximum interference leakage (dBm) versus the minimum required DL SINR (dB), $\Gamma^{\mathrm{DL}}_{\mathrm{req}}$, for different resource allocation schemes.}
\label{fig:system_model}\vspace*{-5mm}
\end{figure}
%%%

For comparison, we consider two baseline resource allocation schemes. For baseline scheme 1, we perform ZF DL transmission for the secondary network where the direction of beamformer $\mathbf{w}_k$ for secondary DL user $k$ is fixed and lies in the null space of the other secondary DL user channels. Then, we jointly optimize $P_j$ and the power of $\mathbf{w}_k$ subject to constraints C1-C4 as in (\ref{eqn:pro}) via SDP relaxation. For baseline scheme 2, we consider a secondary network with an HD BS equipped with $N_\mathrm{T}$ antennas. We set $\log_2(1+\Gamma^{\mathrm{DL}}_{\mathrm{req}_k})=1/2\log_2(1+\Gamma^{\mathrm{DL-HD}}_{\mathrm{req}_k})$ and $\log_2(1+\Gamma^{\mathrm{UL}}_{\mathrm{req}_j})=1/2\log_2(1+\Gamma^{\mathrm{UL-HD}}_{\mathrm{req}_j})$ for a fair comparison. Thus, the required SINRs for the secondary DL and UL users served by the secondary HD BS are $\Gamma^{\mathrm{DL-HD}}_{\mathrm{req}_k}=(1+\Gamma^{\mathrm{DL}}_{\mathrm{req}_k})^2-1$ and $\Gamma^{\mathrm{UL-HD}}_{\mathrm{req}_j}=(1+\Gamma^{\mathrm{UL}}_{\mathrm{req}_j})^2-1$, respectively. Besides, the power consumption of DL and UL transmission for the secondary HD network is divided by two as DL and UL transmission do not overlap. Then, we optimize $\mathbf{w}_k$ and $P_j$ to minimize the maximum interference leakage to the primary users for the optimal MMSE receiver at the secondary HD BS \cite{book:david_wirelss_com}.
%%%%%
It can be observed from Figure 2 that the average maximum interference leakage of the baseline schemes is higher than that of the proposed FD-CR system. In particular, the average maximum interference leakage increases with $\Gamma^{\mathrm{DL}}_{\mathrm{req}}$ for baseline scheme 1
due to the fixed beamforming design. Besides, the average maximum interference leakage of baseline scheme 2 is insensitive to $\Gamma^{\mathrm{DL}}_{\mathrm{req}}$ since the $\mathbf{w}_k$ and $P_j$ are optimized for the considered system setting.

\vspace*{-0mm}
%\subsection{Average Maximum Interference Leakage versus Maximum Channel Estimation Error}
In Figure 3, we study the average maximum interference leakage versus the maximum normalized channel estimation error, $\kappa_{\mathrm{est}}^2$, for a minimum required secondary DL SINR of $\Gamma^{\mathrm{DL}}_{\mathrm{req}}=10$ dB and a minimum required secondary UL SINR of $\Gamma^{\mathrm{UL}}_{\mathrm{req}}=5$ dB. As can be observed, the average maximum interference leakage increases with increasing $\kappa_{\mathrm{est}}^2$. In fact, with increasing imperfectness of the CSI, it is more difficult for the secondary FD BS to perform accurate DL beam-steering.
In particular, more DoF are utilized to reduce interference leakage as the channel uncertainty increases which leads to a higher maximum interference leakage.
Besides, as more DoF are consumed for interference leakage reduction, there are fewer DoF available to suppress the SI which degrades the UL reception in the secondary network. Thus, the secondary UL users are forced to transmit with a higher power to satisfy the UL QoS requirements which in turn results in a larger interference leakage to the primary network.
Furthermore, we note that the baseline schemes cause significantly higher interference leakages compared to the proposed scheme due to their inefficient resource allocation.
\begin{figure}
\centering\vspace*{-0mm}
\includegraphics[width=3.4in]{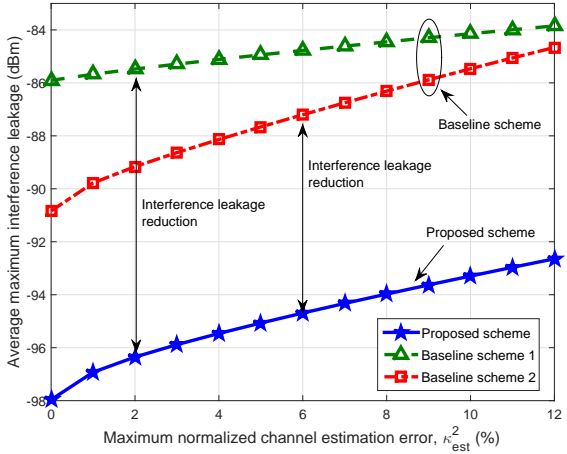}\vspace*{-3mm}
\caption{Average maximum interference leakage (dBm) versus the maximum normalized channel estimation error, $\kappa_{\mathrm{est}}^2$, for $N_{\mathrm{T}}=9$.}
\label{fig:system_model}\vspace*{-6mm}
\end{figure}
\vspace*{-0mm}
\section{Conclusions}
In this paper, we studied the robust resource allocation design for CR secondary networks employing an FD BS for serving multiple secondary HD DL and UL users simultaneously. The algorithm design was formulated
as a non-convex optimization problem with the objective to minimize the maximum interference leakage to the primary network while taking into account the QoS requirements of all secondary users. The imperfectness of the CSI of the secondary network-to-primary network channels was taken into account for robust resource allocation algorithm design. The proposed non-convex problem was solved optimally by SDP relaxation. Simulation results unveiled a significant reduction in interference leakage compared to baseline schemes. Besides, we showed that the proposed scheme is indeed robust with respect to imperfect CSI.

\vspace*{-1mm}
\section*{Appendix - Proof of Theorem 1}
We first solve the convex optimization problem in \eqref{eqn:SDR-robust} and obtain the optimal solution $P_j^{*}$, $\mathbf{W}_k^*$, and the optimal auxiliary variables which are collected in $\mathbf{\Xi}^* \triangleq \{\tau^*, \delta_r^*, \alpha_r^*,\beta_r^*\}$. If $\Rank(\mathbf{W}_k^*)=1,\forall k$, then the globally optimal solution of problem \eqref{eqn:SDR-robust} is achieved. Otherwise, we substitute $P_j^{*}$ and $\mathbf{\Xi}^*$ into the following auxiliary problem: \vspace*{-2mm}
\begin{eqnarray}
\label{eqn:lambda1-0-dlprob}{\underset{\mathbf{W}_k \in \mathbb{H}^{N_{\mathrm{T}}}}{\mino}}\,\, \hspace*{-7mm}&&\hspace*{0mm}\sum_{k=1}^{K}\Tr(\mathbf{W}_k) \notag\\[-2mm]
\mbox{s.t.} \hspace*{0mm}&&\hspace*{0mm}\mbox{C1},\mbox{C2},\mbox{C3},\mbox{C4},\overline{\text{C5}}\mbox{a},\overline{\text{C5}}\mbox{b},\mbox{C6},{\mbox{C8}}.
\end{eqnarray}
Since the problem in (\ref{eqn:lambda1-0-dlprob}) has the same feasible set as problem (\ref{eqn:SDR-robust}), problem (\ref{eqn:lambda1-0-dlprob}) is also feasible. Now, we claim that for a given $P_j^{*}$ and $\mathbf{\Xi}^*$ in (\ref{eqn:lambda1-0-dlprob}), the solution $\mathbf{W}_k^{*}$ of (\ref{eqn:lambda1-0-dlprob}) is a rank-one matrix.
First, the problem in (\ref{eqn:lambda1-0-dlprob}) is jointly convex with respect to the optimization variables and satisfies the Slater's constraint qualification. Therefore, strong duality holds and solving the dual problem is equivalent to solving the primal problem \cite{book:convex}. For obtaining the dual problem, we first need the Lagrangian function of the primal problem in (\ref{eqn:SDR-robust}) which is given by \vspace*{-2mm}
\begin{eqnarray}
\label{eqn:Lagrangian}\notag{\cal L}\hspace*{-1.5mm}&=&\hspace*{-2.5mm} -\sum_{k=1}^{K}\hspace*{-1mm}\lambda_k\hspace*{-1mm}\Tr(\mathbf{H}_k\mathbf{W}_k) \hspace*{-1mm}+\hspace*{-1.5mm}\sum_{j=1}^{J}\hspace*{-1mm}\theta_j\hspace*{-1mm}\sum_{k=1}^{K}\hspace*{-1mm}\Tr(\rho\mathbf{V}_j\hspace*{-1mm}\diag(\mathbf{W}_k\mathbf{H}_{\mathrm{SI}}^H\mathbf{H}_{\mathrm{SI}})) \notag\\[-1mm]
\hspace*{-1.5mm}&+&\hspace*{-2.5mm} (\hspace*{-0.5mm}1\hspace*{-0.5mm}+\hspace*{-0.5mm}\mu\hspace*{-0.5mm})\hspace*{-1mm} \sum_{k=1}^{K} \Tr(\mathbf{W}_k)
-\sum_{r=1}^{R}\Tr(\mathbf{R}_{\overline{\mathrm{C5}}\mathrm{a}_r}\big(\mathbf{W}_k,\alpha_r,\theta_r\big) \mathbf{D}_{\overline{\mathrm{C5}}\mathrm{a}_r})\notag\\[-1mm]
\hspace*{-1.5mm}&-&\hspace*{-2.5mm} \sum_{k=1}^{K}\Tr(\mathbf{W}_k\mathbf{Y}_k)+\Delta.
\end{eqnarray}
Here, $\Delta$ denotes the collection of terms that only involve variables that are independent of $\mathbf{W}_k$. $\lambda_k$, $\theta_j$, and $\mu$ are the Lagrange multipliers associated with constraints ${\mbox{C1}}$, ${\mbox{C2}}$, and ${\mbox{C3}}$, respectively. Matrix $\mathbf{D}_{\overline{\mathrm{C5}}\mathrm{a}_r}\hspace*{-3mm} \in \hspace*{-1mm} {\mathbb{C}^{{(\hspace*{-0.5mm}N_{\mathrm{T}}\hspace*{-0.5mm}+\hspace*{-0.5mm}1\hspace*{-0.5mm})}\times {(\hspace*{-0.5mm}N_{\mathrm{T}}\hspace*{-0.5mm}+\hspace*{-0.5mm}1\hspace*{-0.5mm})}}}$ is the Lagrange multiplier matrix for constraints $\overline{\text{C5}}\mathrm{a}$. Matrix $\mathbf{Y}_k\hspace*{-1mm}\in\hspace*{-1mm}{\mathbb{C}^{N_{\mathrm{T}}\times{N_{\mathrm{T}}}}}$ is the Lagrange multiplier matrix for the positive semidefinite constraint ${\text{C6}}$ on $\mathbf{W}_k$. For notational simplicity, we define $\Psi$ as the set of scalar Lagrange multipliers and $\mathbf{\Phi}$ as the set of matrix Lagrange multipliers.
Thus, the dual problem for the problem in (\ref{eqn:lambda1-0-dlprob}) is given by \vspace*{-2mm}
%\begin{eqnarray}\label{eqn:dual-constraint}
%\underset{\Psi \ge 0, \mathbf{\Phi} \succeq \mathbf{0}}{\maxo} \,\,\underset{\underset{ P_j,\tau,\delta_r, \alpha_{r},\beta_{r}}{\mathbf{W}_k\in\mathbb{H}^{N_{\mathrm{T}}},}}{\mino} \,\,{\cal L} \Big(\hspace*{-0.5mm}\mathbf{W}_k,P_j,\tau,\delta_r, \alpha_{r},\beta_{r},\Psi, \mathbf{\Phi}\hspace*{-0.5mm}\Big).
%\end{eqnarray}
\begin{eqnarray}\label{eqn:dual-constraint}
\underset{\Psi \ge 0, \mathbf{\Phi} \succeq \mathbf{0}}{\maxo} \,\,\underset{\underset{}{\mathbf{W}_k\in\mathbb{H}^{N_{\mathrm{T}}}}}{\mino} \,\,{\cal L} \Big(\hspace*{-0.5mm}\mathbf{W}_k,\Psi, \mathbf{\Phi}\hspace*{-0.5mm}\Big).
\end{eqnarray}
Then, we reveal the structure of the optimal $\mathbf{W}_k$ of (\ref{eqn:lambda1-0-dlprob}) by studying the Karush-Kuhn-Tucker (KKT) conditions.
The KKT conditions for the optimal $\mathbf{W}_k^*$ are given by:\vspace*{-2mm}
\begin{eqnarray}\label{eqn:KKT-multiplier}
\mathbf{Y}_k^*,\mathbf{D}_{\overline{\mathrm{C5}}\mathrm{a}_r}^*
\hspace*{-3mm}&\succeq&\hspace*{-3mm} \mathbf{0},\quad\lambda_k^*, \theta_j^*, \mu^* \ge 0,\\[-0mm]
 \mathbf{Y}_k^*\mathbf{W}_k^*\hspace*{-3mm}&=&\hspace*{-3mm}\mathbf{0}, \label{eqn:KKT-complementarity}\\[-0mm]
\nabla_{\mathbf{W}_k^*}{\cal L}\hspace*{-3mm}&=&\hspace*{-3mm} \mathbf{0}, \label{eqn:subgradient}
\end{eqnarray}
where $\mathbf{Y}_k^*$, $\mathbf{D}_{\overline{\mathrm{C5}}\mathrm{a}_r}^*$, $\lambda_k^*$, $\theta_j^*$, and $\mu^*$ are the optimal Lagrange multipliers for dual problem (\ref{eqn:dual-constraint}), $\nabla_{\mathbf{W}_k^*}{\cal L}$ denotes the gradient of Lagrangian function ${\cal L}$ with respect to matrix $\mathbf{W}_k^*$. The KKT condition in (\ref{eqn:subgradient}) can be expressed as \vspace*{-3mm}
\begin{eqnarray}\label{eqn:KKT-gradient-equivalent}
&&\hspace*{-6mm}(\hspace*{-0mm}1\hspace*{-1mm}+\hspace*{-1mm}\mu^*\hspace*{-0mm})\hspace*{-0.5mm}\mathbf{I}_{N_{\mathrm{T}}} \hspace*{-1mm}+ \hspace*{-1mm} \sum_{j=1}^{J}\hspace*{-1mm}\theta_j^*\rho\mathbf{V}_j\diag(\mathbf{H}_{\mathrm{SI}}^H \mathbf{H}_{\mathrm{SI}}) \hspace*{-1mm}+ \hspace*{-1mm} \sum_{r=1}^{R}\hspace*{-1mm}\mathbf{B}_{\mathbf{l}_r}\hspace*{-0.5mm}\mathbf{D}_{{\overline{\mathrm{C5}}\mathrm{a}}_{r}}^*\hspace*{-2mm} \mathbf{B}_{\mathbf{l}_r}^H \hspace*{-1mm}\notag \\
=&&\hspace*{-6mm}\mathbf{Y}_k^*\hspace*{-1mm} + \hspace*{-1mm}\lambda_k\mathbf{H}_k.
\end{eqnarray}
%Now, we define \vspace*{-2mm}
%%$\mathbf{A}_k^*=\sum_{j=1}^{J}\theta_j^*\mathbf{H}_{\mathrm{SI}}^H \mathbf{V}_j\mathbf{H}_{\mathrm{SI}}+ \sum_{r=1}^{R}\mathbf{B}_{\mathbf{l}_r}\mathbf{D}_{{\overline{\mathrm{C5}}\mathrm{a}}_{r}}^*\mathbf{B}_{\mathbf{l}_r}^H,$ and
%%$\mathbf{\Pi}_k^*=\mu\mathbf{I}_{N_{\mathrm{T}}}+\mathbf{A}_k^*,$
%\begin{eqnarray}\label{eqn:KKT-mess-B}
%\mathbf{\Pi}_k^*=\mu\mathbf{I}_{N_{\mathrm{T}}}+\sum_{j=1}^{J}\theta_j^*\mathbf{H}_{\mathrm{SI}}^H \mathbf{V}_j\mathbf{H}_{\mathrm{SI}}+ \sum_{r=1}^{R}\mathbf{B}_{\mathbf{l}_r}\mathbf{D}_{{\overline{\mathrm{C5}}\mathrm{a}}_{r}}^*\mathbf{B}_{\mathbf{l}_r}^H,
%\end{eqnarray}
%for notational simplicity.
Hence, (\ref{eqn:KKT-gradient-equivalent}) implies \vspace*{-2mm}
\begin{eqnarray}\label{eqn:KKT-shorter}
\mathbf{Y}^*=\mathbf{\Pi}_k^*-\lambda_k\mathbf{H}_k,
\end{eqnarray}
where $\mathbf{\Pi}_k^*\hspace*{-0.5mm}=\hspace*{-0.5mm}(\hspace*{-0mm}1\hspace*{-1mm}+\hspace*{-1mm}\mu^*\hspace*{-0mm})\hspace*{-0.5mm}\mathbf{I}_{N_{\mathrm{T}}} \hspace*{-1mm}+ \hspace*{-2mm} \sum_{j=1}^{J}\hspace*{-1mm}\theta_j^*\rho\mathbf{V}_j\diag(\mathbf{H}_{\mathrm{SI}}^H \mathbf{H}_{\mathrm{SI}}) \hspace*{-1mm}+ \hspace*{-2mm} \sum_{r=1}^{R}\hspace*{-1mm}\mathbf{B}_{\mathbf{l}_r}\hspace*{-0.5mm}\mathbf{D}_{{\overline{\mathrm{C5}}\mathrm{a}}_{r}}^*\hspace*{-2mm} \mathbf{B}_{\mathbf{l}_r}^H \hspace*{-1mm}$.
Premultiplying both sides of (\ref{eqn:KKT-shorter}) by $\mathbf{W}_k^*$, and utilizing (\ref{eqn:KKT-complementarity}), we have \vspace*{-0mm}
$\mathbf{W}_k^*\mathbf{\Pi}_k^* =\lambda_k\mathbf{W}_k^*\mathbf{H}_k$.
%\begin{eqnarray}\label{eqn:KKT-premultiply}
%\mathbf{W}_k^*\mathbf{\Pi}_k^* =\lambda_k\mathbf{W}_k^*\mathbf{H}_k.
%\end{eqnarray}
By applying basic inequalities for the rank of matrices, the following relation holds:\vspace*{-2mm}
\begin{eqnarray}\label{eqn:rank-Wk}
\Rank\big(\mathbf{W}_k^*\big)\hspace*{-1mm}&\overset{(a)}{=}&\hspace*{-1mm}\Rank\big(\mathbf{W}_k^*\mathbf{\Pi}_k^*\big)\notag =\Rank\big(\lambda_k\mathbf{W}_k^*\mathbf{H}_k\big)\notag\\[-2mm]
&\overset{(b)}{\le}&\hspace*{-1mm}\min\Big\{\Rank\big(\lambda_k\mathbf{W}_k^*\big), \Rank\big(\mathbf{H}_k\big)\Big\} \notag\\ [-2mm]
&\overset{(c)}{\le}& \Rank\big(\mathbf{H}_k\big),
\end{eqnarray}
where $(a)$ is due to $\mathbf{\Pi}_k^* \succ \mathbf{0}$, $(b)$ is due to the basic result $\Rank(\mathbf{A}\mathbf{B}) \le \min \big\{ \Rank(\mathbf{A}),\Rank(\mathbf{B})\big\}$, and $(c)$ is due to the fact that $\min\{a,b\} \le a$.
Since $\Rank\big(\mathbf{H}_k\big) \le 1$, by utilizing (\ref{eqn:rank-Wk}), the rank of $\mathbf{W}_k^*$ is given by \vspace*{-1mm}
\begin{eqnarray}\label{eqn:rank-Wk-inequality}
\Rank(\mathbf{W}_k^*)&\le& \Rank\big(\mathbf{H}_k\big) \le 1.
\end{eqnarray}
We note that $\mathbf{W}_k^* \neq \mathbf{0}$ for $\Gamma^{\mathrm{DL}}_{\mathrm{req}_k} > 0$. Thus, $\Rank(\mathbf{W}_k^*)=1$. Therefore, an optimal rank-one matrix $\mathbf{W}_k^{*}$ for \eqref{eqn:SDR-robust} is constructed.

%\bibliographystyle{IEEEtran}
%\bibliography{FD_cognitive}

\end{document}